\documentclass[aps,prl,twocolumn,showpacs, amssymb, amsmath, aps, superscriptaddress] {revtex4-1}

\usepackage{graphicx}
\usepackage{dcolumn}
\usepackage{bm}
\usepackage{epstopdf}
\usepackage{latexsym}
\usepackage{mathrsfs}
\begin{document}

\title{Small, Highly Accurate Quantum Processor for Intermediate-Depth Quantum Simulations}

\author{Nathan K. Lysne}\affiliation{Center for Quantum Information and Control, Wyant College of Optical Sciences, University of Arizona, Tucson, AZ 85721, USA}
\author{Kevin W. Kuper}\affiliation{Center for Quantum Information and Control, Wyant College of Optical Sciences, University of Arizona, Tucson, AZ 85721, USA}
\author{Pablo M. Poggi}\affiliation{Center for Quantum Information and Control, Department of Physics and Astronomy, University of New Mexico, Albuquerque, NM 87131, USA}
\author{Ivan H. Deutsch}\affiliation{Center for Quantum Information and Control, Department of Physics and Astronomy, University of New Mexico, Albuquerque, NM 87131, USA}
\author{Poul S. Jessen}\affiliation{Center for Quantum Information and Control, Wyant College of Optical Sciences, University of Arizona, Tucson, AZ 85721, USA}

\date{\today}

\begin{abstract}
Analog quantum simulation is widely considered a step on the path to fault tolerant quantum computation. With current noisy hardware, the accuracy of an analog simulator will degrade after just a few time steps, especially when simulating complex systems likely to exhibit quantum chaos. Here we describe a quantum simulator based on the combined electron-nuclear spins of individual Cs atoms, and its use to run high fidelity simulations of three different model Hamiltonians for ${>}100$ time steps. While not scalable to exponentially large Hilbert spaces, it provides the accuracy and programmability required to explore the interplay between dynamics, imperfections, and accuracy in quantum simulation. 
 \end{abstract}
                              
\maketitle

Absent errors, machines that process information according to quantum mechanics can in principle solve problems beyond the computational power of any classical computer. In practice, a scalable, general purpose quantum computer must include error correction and fault tolerance as an integral part of its operation, leading to requirements on the underlying quantum hardware that could be out of reach for years to come \cite{Campbell2017}. Thus, in the current era of noisy, intermediate-scale quantum (NISQ) devices \cite{Preskill2018}, much of the effort in the field has focused on seemingly less ambitious challenges. High on the list is the development of analog quantum simulators, defined here as devices that operate without error correction but still have the potential to surpass classical computers for tasks such as modeling complex quantum systems \cite{Georgescu2014, Tacchino2019}. Recent examples include work using traped ions \cite{Zhang2017, Hempel2018, SafaviNaini2018}, Rydberg atoms \cite{Labuhn2016, Bernien2017}, and superconducting qubits \cite{Barends2016, Harris2018} to simulate phase transitions and other phenomena in large (${>}50$) spin systems. This is roughly the scale at which numerical modeling on classical computers is currently infeasible. 

Quantum simulation generally requires access to highly entangled states of interacting many body systems. It has long been known that such systems also tend to support quantum chaos, in the sense that their time evolution is hypersensitive to perturbation \cite{Peres1991, Georgeot2000, Georgeot2007}. This suggests two separate notions of complexity relevant for quantum simulation, one related to the nature of the quantum state, and another related to the nature of the system dynamics. Entangled states are complex because the information required to predict interparticle correlations grows exponentially with system size, while chaotic dynamics are complex because the information required to predict the quantum trajectory grows exponentially with time \cite{Schack1997}.  Both will contribute to the overall complexity and fragility of analog quantum simulation and related NISQ-era objectives such as quantum annealing \cite{Albash2019, Pearson2019}. Indeed, one can expect an inverse relationship between the accessible Hilbert space and the length of time one can meaningfully simulate, with those properties playing a role analogous to the width and depth in quantifying the complexity of a quantum circuit. So far, experiments have focused mostly on the ÓwidthÓ of a simulation (after all, this is the crucial resource when looking for a quantum advantage), with limited attention paid to the fidelity of the output state as one seeks to increase its ÓdepthÓ in terms of simulated time. Yet, to fully understand the computational power of an analog quantum simulation, it is necessary to look carefully at the accessible simulation depth and how it depends on the nature of the dynamics, before one can trust its outcome \cite{Hauke2012}. 

In this letter we present a new platform for analog quantum simulation (AQS) with tradeoffs that are complementary to NISQ devices: it is modest in terms of accessible Hilbert space, but highly accurate and therefore uniquely suited to the study of dynamical complexity in time. Our small, highly accurate quantum (SHAQ) simulator is based on the combined electron-nuclear spins of individual Cs atoms in the electronic ground state, driven by phase modulated radio-frequency (rf) and microwave ($\mu$w) magnetic fields, and provides access to a fixed $16$-dimensional Hilbert space formally equivalent to four qubits. In place of the quantum circuit model where control is predicated on access to a universal gate set, we rely on a universal control Hamiltonian and quantum Optimal Control \cite{Merkel2008, Smith2013} to ensure that our simulator is fully programmable, in the sense that we can implement arbitrary unitary maps with average fidelities ${>}0.98$ \cite{Anderson2015}. We show that Optimal Control can be further adapted for AQS, allowing us to set up coarse-grained simulations of the dynamics driven by arbitrary model Hamiltonians. Thus, while Optimal Control does not scale to exponentially large Hilbert spaces, it delivers a combination of flexibility and accuracy that is uniquely suited for AQS on SHAQ hardware. We demonstrate experimental AQS of three different spin Hamiltonians on the same device, for ${>}100$ time steps with average fidelity ${>}0.99$ per step. Our work stands out by directly measuring the fidelity of the evolving quantum state as a simulation proceeds. This is the ultimate metric for accuracy, yet quantum state fidelity is rarely reported in contemporary AQS experiments with NISQ hardware, perhaps because indirect measures based on quantum state tomography involve complex protocols that are prone to their own errors \cite{SosaMartinez2017}.  For examples of other ways to estimate fidelity without resorting to full quantum state tomography, see  \cite{Flammia2011, daSilva2011, Lanyon2017, Elben2020}. More conventionally, we demonstrate experimental AQS of the time evolution of a ÓbulkÓ observable (magnetization), with a simulation depth and accuracy that lies well beyond the capabilities of universal simulators based on current NISQ hardware (see, for example, refs. \cite{Lanyon2011, Chiesa2019, Smith2019, Gustafson2019}). These are key elements required for future exploration of the interplay between dynamics, hardware imperfections, and accuracy in AQS.

\begin{figure}
\includegraphics[width=87mm]{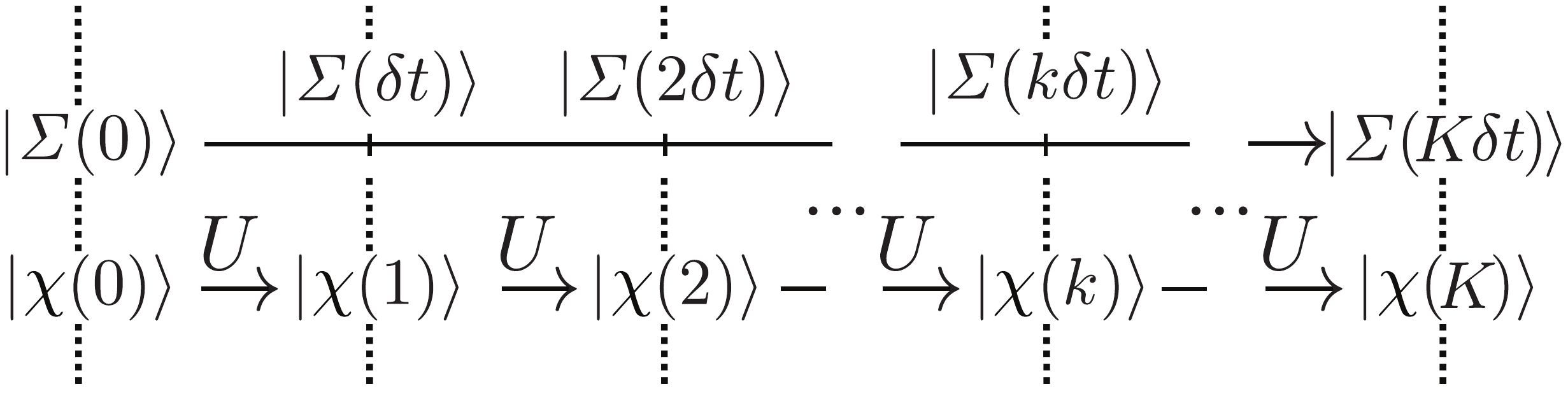}
\caption{Analog Quantum Simulation.  Top: The system evolves continuously from the initial state $| {\it\Sigma}(0) \rangle$ to the final state $| {\it\Sigma}(K \delta t) \rangle$ according to the system Hamiltonian. Bottom: The simulator evolves from the initial state $| \chi(0) \rangle$ to the final state $| \chi(K) \rangle$, coinciding with the system state at intervals $k{=}0,1,2,...,K$.}
\label{fig:fig1}
\end{figure}     

AQS falls into two broad categories sometimes referred to as ``emulation'' and ``simulation''. A quantum emulator is a special-purpose device governed by the same Hamiltonian and having the same Hilbert space structure as the system of interest. Emulators have been realized on a variety of physical platforms and used to study a range of phenomena with considerable success \cite{Zhang2017, Britton2012, Jurcevic2014, Gross2017}. A quantum simulator, by contrast, is a universal device that is controllable, in the sense that one can ``program'' it to implement any $\text{SU}(d)$ map of Hilbert space onto itself. Given an arbitrary system Hamiltonian and a mapping of the system Hilbert space $\mathscr{H}_{\text{sys}}$ onto the simulator Hilbert space $\mathscr{H}_{\text{sim}}$, one can then implement unitary time steps on the simulator and iterate to perform stroboscopic simulations of the system dynamics (Fig. 1). 

Our Cs-atom based ``device'' is a quantum simulator in the second sense, and requires unitary control over the entire accessible Hilbert space. As shown in \cite{Merkel2008}, the spin degrees of freedom of a Cs atom in its electronic ground state are controllable with a combination of phase modulated rf and $\mu$w magnetic fields whose piecewise constant phases $\{\phi_{i}^{\text{rf}x}{,}\phi_{i}^{\text{rf}y}{,}\phi_{i}^{\mu \text w}\}{=}\{\vec{\phi}_i\}$, $1 {\le} i {\le} N_\phi$, serve as control variables (``controls'' for short). We can then apply the generic toolbox of quantum Optimal Control to find (non-unique) controls that accomplish the control task at hand.  In this article we focus on the adaptation of Optimal Control to AQS, and refer the reader to past work \cite{Merkel2008, Smith2013, Anderson2015, SosaMartinez2017} and \cite{SupplementalMaterial} for details of the laboratory implementation. 

\begin{figure*}
\includegraphics[width=18cm]{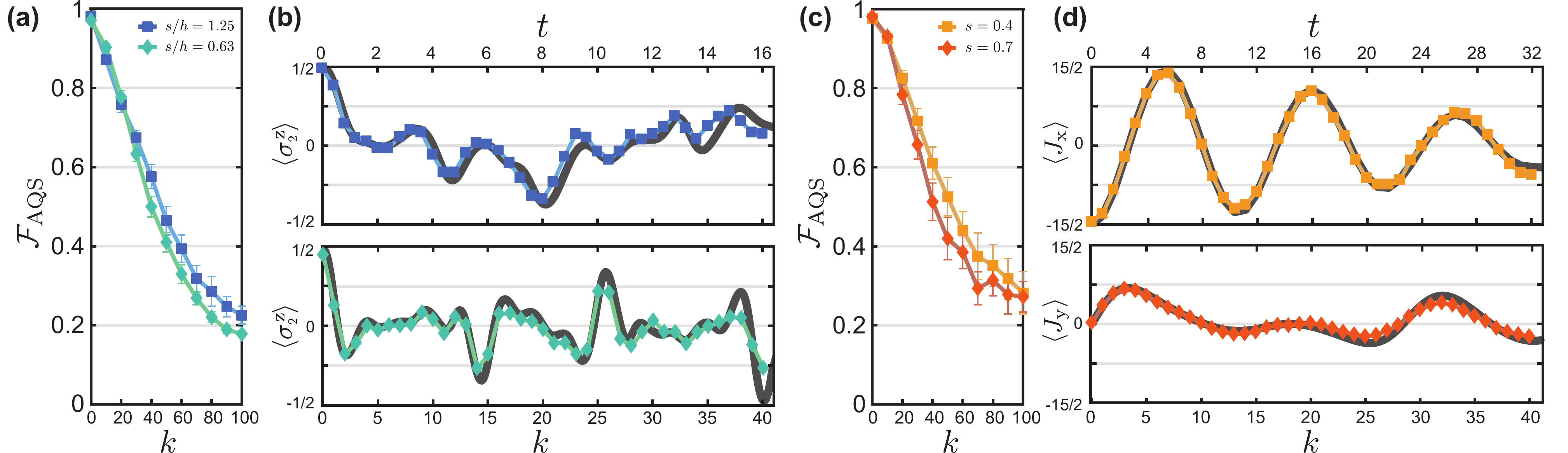}
\caption{(Color online) AQS of two popular model systems corresponding to the Hamiltonians in Eqs. 2a,b.  (a) Measured $\mathcal{F}_\mathrm{AQS}(k)$ for two versions of the TI model with $\delta t{=}0.4, s/h{=}1.25$ (squares) and $\delta t{=}0.4, s/h{=}0.63$ (diamonds). (b) Simulation results for $\langle \sigma_2^z \rangle$ at $\delta t{=}0.4, s/h{=}1.25$ (top) and $\delta t{=}0.4, s/h{=}0.63$   (bottom). (c) Measured $\mathcal{F}_\mathrm{AQS}(k)$ for two versions of the LMG model with $\delta t{=}0.8,s{=}0.4$ (squares) and  $\delta t{=}1.6,s{=}0.7$ (diamonds).  (d) Simulation results for $\langle J_x \rangle$ at $\delta t{=}0.8,s{=}0.4$ (top) and  $\langle J_y \rangle$ at $\delta t{=}0.8,s{=}0.7$ (bottom). Thin lines connect the data points to guide the eye. Thick solid lines in (b) and (d) shows predictions of the exact model.  Error bars are the standard error of the mean for the chosen sample of initial states.}
\label{fig:fig2}
\end{figure*} 

To set up an AQS we first choose orthonormal bases in $\mathscr{H}_{\text{sys}}$ and $\mathscr{H}_{\text{sim}}$. Having done so, a straightforward way of mapping from system to simulator is to represent states and operators by identical vectors and matrices in the two bases. Next, given a unitary time propagator $W$ acting on the system, we seek controls $\{\vec{\phi}_{i}\}$  for which the transformation $U(\{\vec{\phi}_{i}\})$ acting on the simulator is a good approximation to $W$. Note that $W$ can be chosen as the exact propagator for a time step of any length; there will be no Trotter errors \cite{Georgescu2014, Tacchino2019} unless deliberately introduced as part of the simulation. Given $W$, we then use one of two versions of Optimal Control to find high-performing controls:

\noindent \underline{Conventional Control.} This version uses an objective function $\mathcal{F}(\{\vec{\phi}_{i}\}){=}|\text{Tr}[W^{\dagger}U(\{\vec{\phi}_{i}\})]|^2/{d^2}$  (the fidelity). Numerical optimization of $\mathcal{F}(\{\vec{\phi}_{i}\})$ as a function of $\{\vec{\phi}_{i}\}$ will find controls for which the matrices $W$ and $U(\{\vec{\phi}_{i}\})$ are near identical in the chosen bases. That is, within a global phase, they have the same eigenvalues and eigenstates. In prior work we have found that $N_\phi{=}150$ phase steps (450 phase values) is sufficient to consistently achieve a theoretical $\mathcal{F}(\{\vec{\phi}_{i}\}){\ge} 0.99999$  for any $W$. In the laboratory where control errors and decoherence are present, the corresponding controls achieve $\mathcal{F}{\ge} 0.98$ on average, as measured by randomized benchmarking \cite{Anderson2015}.

\noindent \underline{EigenValue Only (EVO) Control.} In AQS there are in principle no restrictions on the map from system to simulator. To take advantage of this, we note that it suffices for $W$ and $U(\{\vec{\phi}_{i}\})$ to have near identical eigenvalues, in which case $W$ and $V U(\{\vec{\phi}_{i}\}) V^\dagger$ will be nearly identical for some $V$. Accordingly, the EVO approach uses an objective function 
\begin{equation}
\mathcal{F}_{\mathrm{EVO}}(\{\vec{\phi}_{i}\}{,}\{v_j\}){=}\frac{1}{d^{2}} | \text{Tr} [W^{\dagger} V U(\{\vec{\phi}_{i}\}) V^{\dagger}] |^{2}
\end{equation}

\noindent where $V{=}e^{iA}$, $A{=}\sum_{j=1}^{d^2-1}v_j \Lambda_j$, $\{\Lambda_j\}$ is a set of generalized Gell-Mann matrices forming a basis of traceless hermitian $d{\times} d$ matrices, and $\{v_j\}$ is a set of $d^2{-}1$ real-valued variables, sufficient to generate all $V{\in} \text{ SU}(16)$. Simultaneous optimization of $\mathcal{F}_{\mathrm{EVO}}(\{\vec{\phi}_{i}\}{,}\{v_j\})$ with respect to $\{\vec{\phi}_{i}\}$ and $\{v_j\}$ will then find co-optimal controls and system-simulator maps.  

Our experience suggests the search complexity and computational effort is comparable for Conventional and EVO Control. Furthermore, we find that optimizing for EVO rather than the entire $W$ can reduce the number of phase steps from $N_{\phi}{=}150$ to something in the range from $N_{\phi}{=}10$ to $N_{\phi}{=}60$, depending on the nature of $W$. This brings a significant advantage in terms of the possible number of time steps and overall fidelity in an AQS. On the downside we have found that EVO Control performs poorly when $W$ is close to the identity. This is generally not an issue for AQS where one typically chooses time steps that change the state appreciably. A second issue arises if the system Hamiltonian is time dependent and the propagators $W(k)$ are different for different time steps $k$. One must then do separate EVO searches for solutions $V_k U_k(\{\vec{\phi}_{i}\}) V_k^{\dagger}$, where inevitably $V_k {\ne} V_{k+1}$. The resulting basis mismatch means one cannot simply concatenate the $U_k(\{\vec{\phi}_{i}\})$, and any attempt to restrict the $V_k$'s negates the original advantage. Therefore, when necessary, we revert to Conventional Control which works in every scenario we have explored.  See \cite{SupplementalMaterial} for details.

To establish the baseline performance of our quantum simulator we have tested it on three popular model systems described by the Hamiltonians 

\begin{align}
&H_{\text{TI}}={-}\sum_{i{=}1}^N h\sigma_i^z{-}\sum_{i{=}1}^{N{-}1} s\sigma_i^x \sigma_{i{+}1}^x,\quad {N=4} \tag{2a} \\
&H_{\text{LMG}}={-}(1{-}s)J_z{-}sJ_x^2, \qquad   J=15/2 \tag{2b}\\
&H_{\text{QKT}}={-}p J_z \sum_{n{=}0}^\infty \delta(\tau{-}n T){-}\frac{\kappa}{2JT} J_x^2, \quad J=15/2 \tag{2c}\\
\nonumber
\end{align}

\noindent The nearest-neighbor Transverse Ising (TI) \cite{Suzuki2013} and Lipkin-Meshkov-Glick (LMG) \cite{Lipkin1965, Zibold2010} models are common paradigms for the study of phase transitions, and while integrable they nevertheless feature nontrivial dynamics. The Quantum Kicked Top (QKT) is a time-discrete version of the LMG model whose classical phase space can be regular, mixed or globally chaotic depending on the parameters $p, \kappa$  \cite{Haake1987, Chaudhury2009}. For each model we choose the system size $N$ or $J$ to use the entire $16$-dimensional Hilbert space available on our simulator.  

In the laboratory each AQS follows the same basic template. Given a model Hamiltonian, we use Conventional or EVO Control to find controls $\{\vec{\phi}_{i}\}$ and a corresponding propagator $U(\{\vec{\phi}_{i}\})$ that simulates the system evolution during a time step $\delta t$. Knowing the system-simulator map, we then prepare the simulator in the chosen initial state $|\chi(0)\rangle$, take $k$ time steps, measure the observable $M$ of interest, and repeat for $1{\leq} k {\leq} K$ to build up a stroboscopic record of the expectation value $\langle M(k) \rangle$.  In AQS of systems such as the TI and LMG models, $M$ might be a spin observable. As a measure of the accuracy of the simulation we can also look at the fidelity of the quantum state itself, $\mathcal{F}_\mathrm{AQS}(k){=}\text{Tr}[\rho_{\text{a}}(k)|\chi(k) \rangle \langle \chi(k) |] $, where $|\chi(k) \rangle$ and $\rho_{\text{a}}(k)$ are the predicted and actual states after $k$ steps. To access this quantity we measure the projector $M(k){=}|\chi(k) \rangle \langle \chi(k) |$, i. e., the probability of finding the simulator in the predicted state after $k$ steps. In practice, using laser cooling to prepare a large sample of non-interacting Cs atoms allows us to run as many as $10^7$  identical quantum simulators in parallel. This leads to small variations in the control fields from atom to atom, but ensures excellent averaging over noise in the controls and the measurement.  In our setup the time per phase step is $4\mu$s and the maximum overall duration of a quantum simulation is $12$ms, limited by the time a free-falling atom spends in the region of uniform control fields.

\begin{figure*}
\includegraphics[width=18cm]{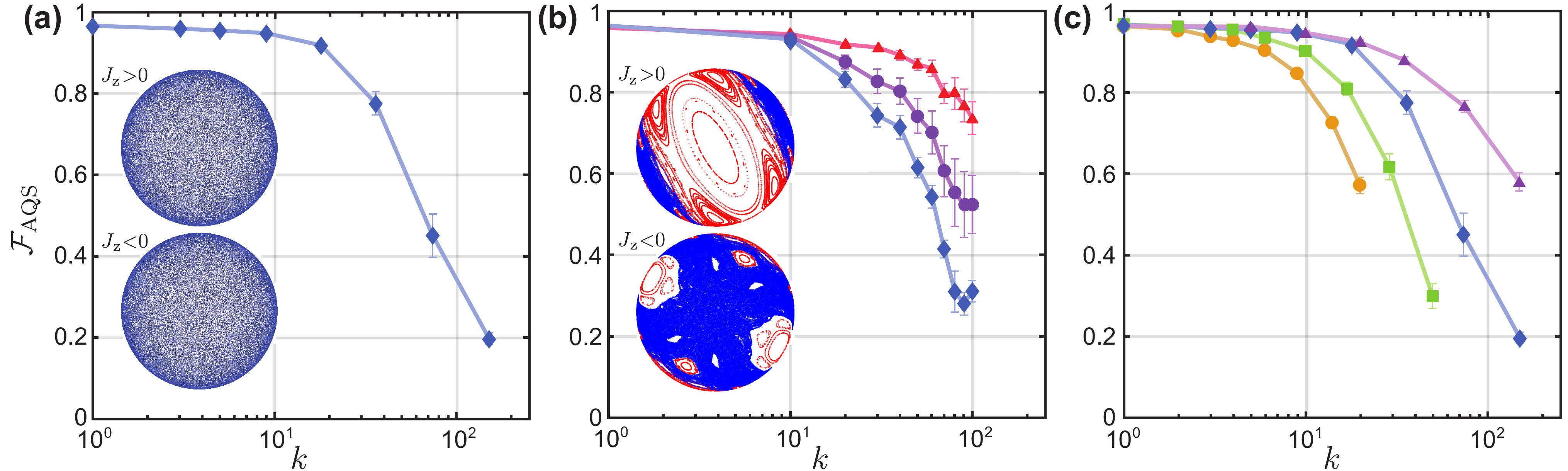}
\caption{(Color online) AQS of the Quantum Kicked Top, Eq. 2c.  (a) Measured $\mathcal{F}_\mathrm{AQS}(k)$, averaged over 10 initial coherent states, in the globally chaotic regime $(p{=}1{,}\kappa{=}7)$ .  Insert: Classical phase space map for the hemispheres $J_z{>}0$ and $J_z{<}0$ showing global chaos. (b) Measured $\mathcal{F}_\mathrm{AQS}(k)$ in a mixed regime $(p{=}0.99,\kappa{=}2.3)$, averaged over $5$ initial coherent states in the large regular island in the $J_z{>}0$ hemisphere (triangles), over $5$ initial coherent states in the sea of chaos in the $J_z{<}0$ hemisphere (diamonds), and over all $10$ initial states (circles). Insert: Classical phase space map. (c) Measured $\mathcal{F}_\mathrm{AQS}(k)$ ($p{=}1{,}\kappa{=}7)$, averaged over $10$ initial coherent states, for different control strategies: Conventional Control with $N_\phi{=}150,K{=}20$ (circles), EVO Control with $N_\phi{=}60,K{=}50$ (squares) and $N_\phi{=}20,K{=}150$ (diamonds), and finally randomized EVO Control with $N_\phi{=}20,K{=}150$ where coherent errors are scrambled (triangles). Thin lines connect the data points to guide the eye. Error bars are the standard error of the mean for the chosen sample of initial states}
\label{fig:fig3}
\end{figure*} 

Figures 2a,c show $\mathcal{F}_\mathrm{AQS}(k)$  for two versions of a $4$-site TI model and a $J{=}15/2$ LMG model, each for $0{\leq}k{\leq}100$ time steps. The simulations use EVO Control with $N_\phi{=}20$ phase steps, the fidelities are averages over $10$ randomly chosen initial states, and the cut-off at $K{=}100$ is to ensure the fidelity of the time steps does not vary with $k$. The TI data corresponds to ferromagnetic and paramagnetic regimes respectively, while the LMG data correspond to regimes below and just above the critical point. In all cases the simulation fidelity declines smoothly with time but remains well above that of a mixed state, $\mathcal{F}_\mathrm{AQS}(k){>}1/16{=}0.0625$, even after $k{=}100$ time steps. As examples of AQS of dynamical observables, Figs. 2b,d show simulations of $\langle \sigma_2^z \rangle$ for the TI model, and collective spin components $\langle J_x \rangle$ and $\langle J_y \rangle$ for the LMG model; these examples were chosen because of their varied and nontrivial behavior. Each AQS extends over $1{\leq}k{\leq}40$ time steps, enough to densely sample for long enough that the nature of the dynamics becomes apparent.  Overall, quantum simulation of the TI model appears somewhat more challenging than the LMG model, but both track the spin dynamics with good accuracy.  Notably, $40$ time steps is enough for a considerable loss of fidelity, $\mathcal{F}_\mathrm{AQS}(k){ \sim}0.5$, showing that useful simulation of physical observables can be achieved even when accuracy at the level of the quantum state is fairly poor.

As a third test we apply our simulator to the Quantum Kicked Top. An AQS of the QKT consists of repeated applications of the Floquet Operator, and the only meaningful time step is one period of $H_\mathrm{QKT}.$ This assures that the propagator $W$ is far from the identity for all but a small subset of parameters, and thus ideally suited for EVO Control.  Figs. 3a,b show fidelities $\mathcal{F}_\mathrm{AQS}(k)$ for two versions of the QKT with globally chaotic $(p{=}1, \kappa{=}7 )$ and mixed $(p{=}0.99, \kappa{=}2.3 )$ phase spaces, averaged over $10$ initial coherent states of the collective spin. Based on general arguments relating quantum chaos to hypersensitivity \cite{Peres1991}, one might expect that AQS will be more challenging for the globally chaotic case than for the mixed case, and this may explain the lower fidelity seen for the former.  For the mixed case we can separate out initial states in regular versus chaotic regions, with the former achieving significantly higher fidelity than the latter. Somewhat counter to expectations, however, every AQS of the QKT shown here does at least as well, and in some case significantly better in terms of fidelity per step, than AQS of the integrable TI and LMG models. 

Finally, we use the QKT in the globally chaotic regime to explore how the fidelity of an AQS depends on the control strategy and the errors present in the experiment. Figure 3c shows simulation fidelities for three scenarios: EVO Control with $N_\phi{=}20$ and $N_\phi{=}60$ phase steps, and Conventional Control with $N_\phi{=}150$; the corresponding maximum number of QKT steps are $K{=}150$, $K{=}50$, and $K{=}20$, and the fidelities decline smoothly to $\mathcal{F}_\mathrm{AQS}(k){\sim}0.2$, $\mathcal{F}_\mathrm{AQS}(k){\sim}0.3$, and $\mathcal{F}_\mathrm{AQS}(k){\sim}0.58$ at the end points. To our knowledge, this demonstrates an accessible simulation depth that compares favorably with current state of the art for AQS on NISQ hardware.

It is worth recalling that every AQS discussed here involves repeated application of the time propagator over and over again, a scenario in which coherent control errors have the potential to compound much faster than random noise. We can explore the role of coherent errors versus noise by comparing to a scenario akin to randomized benchmarking \cite{Knill2008}. To do so, we use EVO optimization with $N_\phi{=}20$ to find a number of time propagators $U(\{\vec{\phi}_{i}\})$, all with identical eigenvalues but different and effectively random controls $\{\vec{\phi}_{i}\}$ and maps $V$.  These are put together in random sequences of various lengths, at the end of which we measure the fidelity of the output state relative to the output predicted in the absence of errors.  As seen in Figure 3c, the resulting decline in fidelity is significantly slower than for each of the three quantum simulations, with an end point fidelity of $\mathcal{F}_\mathrm{AQS}(k){\sim}0.6$ at $k{=}150$, and an average fidelity per step of $0.997$.  We take this as evidence that quantum simulations such as those studied here are more strongly affected by coherent errors than random noise, and that efforts to improve the simulation fidelity should focus on the former. 

In conclusion, we have demonstrated a small, universal and highly accurate analog quantum simulator based on the spin-degrees of freedom in the ground state of individual Cs atoms. We have further shown how Optimal Control can be adapted to program such a simulator, and we have established its baseline performance by applying it to both integrable and chaotic model systems. Notably, the idea of looking for co-optimal controls and system-simulator maps is not restricted to quantum simulation, and could lead to similar gains in other contexts where a generic control task is mapped onto a specific piece of hardware. Going forward, we plan to use our Cs atom quantum simulator to develop and test a general model of the interplay between the native errors of a generic quantum simulator and the types of observables one might use it to access.  Ultimately, we hope testbeds such as ours can help better understand the computational power of analog quantum devices and their use in lieu of error corrected and fault tolerant quantum computers.

\begin{acknowledgments}
The authors are grateful to Prof. Christiane Koch at Freie Universit\"at Belin and to Anupam Mitra at the University of New Mexico for helpful discussions.  This work was supported by the US National Science Foundation Grants No. 1521439, 1820679, and 1630114. 
\end{acknowledgments}


\end{document}


\title{Supplemental Material:\\
A Small, Highly Accurate Quantum Processor for Intermediate-Depth Quantum Simulations}

\author{Nathan K. Lysne}\affiliation{Center for Quantum Information and Control, Wyant College of Optical Sciences, University of Arizona, Tucson, AZ 85721, USA}
\author{Kevin W. Kuper}\affiliation{Center for Quantum Information and Control, Wyant College of Optical Sciences, University of Arizona, Tucson, AZ 85721, USA}
\author{Pablo M. Poggi}\affiliation{Center for Quantum Information and Control, Department of Physics and Astronomy, University of New Mexico, Albuquerque, NM 87131, USA}
\author{Ivan H. Deutsch}\affiliation{Center for Quantum Information and Control, Department of Physics and Astronomy, University of New Mexico, Albuquerque, NM 87131, USA}
\author{Poul S. Jessen}\affiliation{Center for Quantum Information and Control, Wyant College of Optical Sciences, University of Arizona, Tucson, AZ 85721, USA}

\date{\today}

\maketitle

\section{Basics of Control and Measurement}
\noindent The structure of the $6S_{1/2}$ electronic ground state of $^{133}$Cs follows from the addition of electron and nuclear spins. The resulting Hilbert space has two manifolds with quantum numbers $F^{(\pm)}=I \pm S=3,4$, and a total of 16 magnetic sublevels $|F,m \rangle$ (Fig. 1). As shown in \cite{Merkel2008, Smith2013}, this system is controllable with a static bias magnetic field along $z$, a pair of phase-modulated rf magnetic fields along $x$ and
$y$, and a phase-modulated $\mu\text{w}$ magnetic field coupling states $|F^{(\pm)}, m=F^{(\pm)} \rangle$.  In the rotating wave approximation, the
control Hamiltonian has the folllowing structure, 
\begin{equation*}
\begin{split}
H_{\text{c}}(t) = H_{0} + H_{\text{rf}}^{(+)} [ \phi_{x}(t), \phi_{y}(t) ] + H_{\text{rf}}^{(-)} [ \phi_{x}(t), \phi_{y}(t) ] \\ + H_{\mu\text{w}} [ \phi_{\mu\text{w}}(t) ]
\end{split}
\end{equation*}

\noindent Here $H_{0}$ is a drift term including the hyperfine interaction and Zeeman shift from the bias field, the $H_{\text{rf}}^{(\pm)}$ generate $SU(2)$ rotations of the $F^{(\pm)}$ hyperfine spin manifolds depending on the rf phases, and $H_{\mu\text{w}}$ generates $SU(2)$ rotations of the $|F^{(\pm)},m=F^{(\pm)} \rangle$ pseudospin depending on the $\mu\text{w}$ phase. Besides the control phases, $H_{\text{c}}(t)$ depends on the following parameters (nominal values in parenthesis): the Larmor frequency in the bias field ($\Omega_{0} = 2\pi \times 1\text{MHz}$), the rf Larmor frequencies in the rotating frame ($\Omega_{x} = \Omega_{y} = 2\pi \times 25 \text{kHz}$), the $\mu\text{w}$ Rabi frequency ($\Omega_{\mu\text{w}} = 2 \pi \times 27.5 \text{kHz}$), and the rf and $\mu\text{w}$ detunings from resonance ($\Delta_{\text{rf}} = \Delta_{\mu\text{w}} = 0$).  For these parameter values we have found empirically that a phase step duration of $4\mu$s is close to optimal \cite{Anderson2015}.

\begin{figure}
\includegraphics[width=85mm]{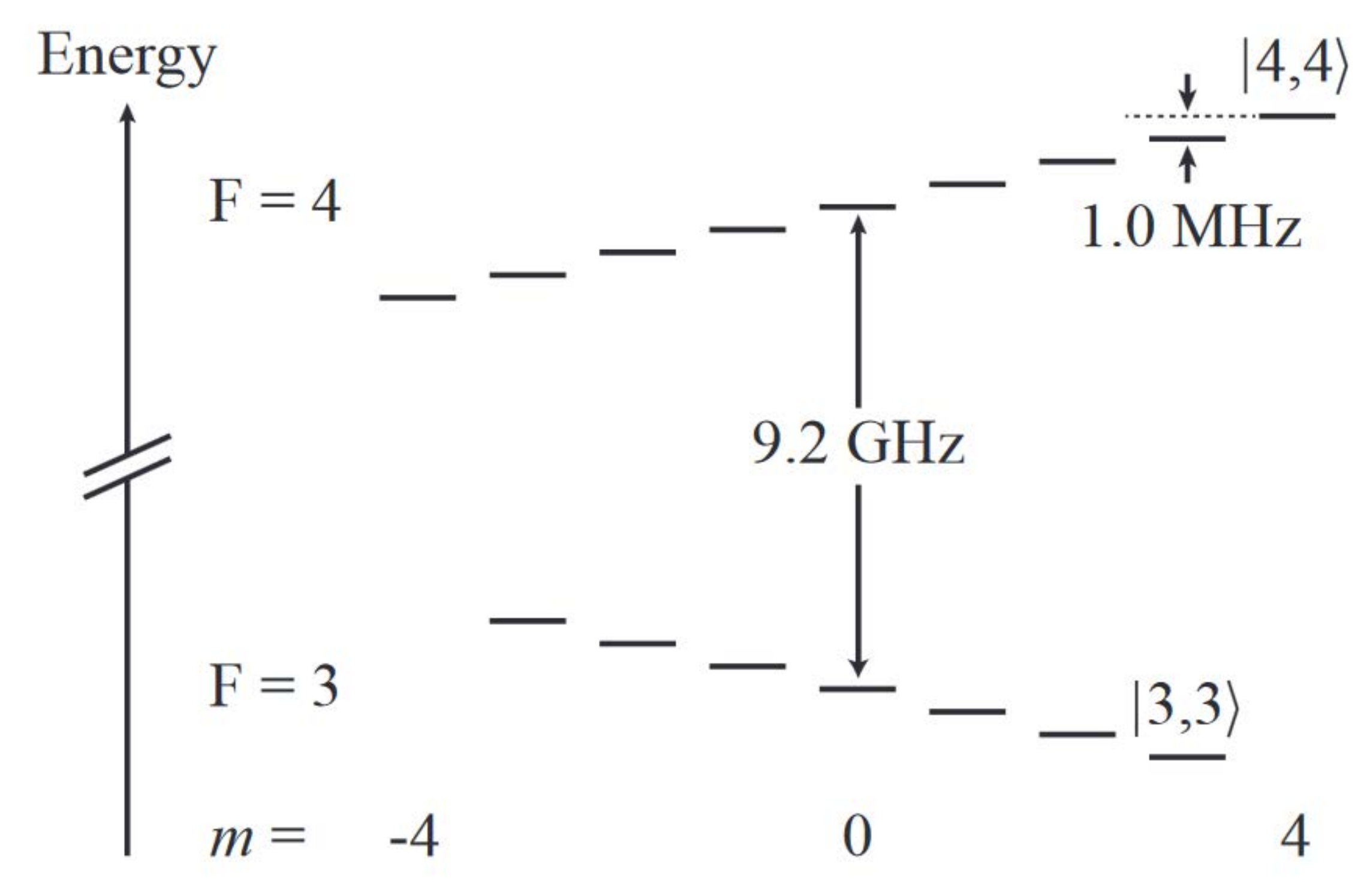}
\caption{Hyperfine structure in the $^{133}$Cs $6S_{1/2}$ electronic ground state. A bias magnetic field of ${\sim}$3 Gauss, corresponding to a Larmor frequency of 1.0 MHz, is applied to remove degeneracy between magnetic sublevels.}
\end{figure} 
        
A typical experimental sequence begins with an ensemble of $\sim 10^{7}$  laser cooled atoms released from a magneto-optic trap/optical molasses into free fall. We use optical pumping to prepare the atoms in $|\psi_{0} \rangle = |F=3,m=3 \rangle$, and then use Optimal Control to implement a state map $|\psi_{0} \rangle \rightarrow |\chi(0) \rangle = \sum_{F,m} c_{F,m} |F,m\rangle$ to prepare the desired initial state. In practice errors and imperfections in the preparation sequence cause the actual state, $\rho_{\text{a}}$, to deviate slightly from the intended target, with a typical infidelity $1-\langle \chi_{0} | \rho_{\text{a}} | \chi_{0} \rangle\approx0.5\%$. 

The basic resource for measurement in our experiment is Stern-Gerlach analysis, implemented by letting the atoms fall in a magnetic field gradient, and collecting the time dependent fluorescence as they pass though a probe beam approximately $5\text{cm}$ below the location of the magneto-optic trap. The resulting signal is fitted to extract the measurement probabilities, $p_{F,m} = \langle F,m|\rho_{\text{a}}|F,m\rangle$, with fluctuations in probe power and detector electronic noise leading to a roughly $1\%$ uncertainty in these estimates. With the basic Stern-Gerlach measurement in place, our ability to perform $SU(16)$ maps makes it straightforward to implement additional POVMs. Specifically, we can perform a 16-outcome measurement in an arbitrary basis $\{|\Psi_{\alpha}\rangle\}$, by mapping each $\{|\Psi_{\alpha}\rangle\}$ onto a magnetic sublevel $|F_{\alpha},m_{\alpha}\rangle$ through a unitary transformation $ U =\sum_{\alpha} |F_{\alpha},m_{\alpha}\rangle \langle \Psi_{\alpha}|$. We then fit the resulting Stern-Gerlach analysis signals to find the outcome probablities $\{p_{\alpha}\}$ \cite{SosaMartinez2017}.

One of the novel and powerful features of our simulator is the ready access to direct measurements of quantum state fidelity, without resorting to complex and time consuming diagnostics such as quantum tomography.  Even the most complex system states, such as those resulting from time evolution under the exact TI, LMG, and QKT Hamiltonians, can be calculated in advance.  A set of controls can then be designed to map the equivalent state on the simulator to a convenient hyperfine state, say, $|F=4,m=4\rangle$, with the orthogonal complement mapped to the space spanned by the remaining $15$ hyperfine states. In the experiment we apply this "read-out" map at the end of the simulation sequence, and then use Stern-Gerlach analysis to determine the probability of having the simulator emerge in the correct final state, now $|F=4,m=4\rangle$.  This is, by definition, the fidelity.  Based on randomized benchmarking \cite{Anderson2015}, we estimate that the accuracy of this measure is within 2-3\%. 

Finally, it is worth noting that the fidelity as we measure it here is closely related to the Loschmidt echo \cite{Gorin2006}.  We can see this by rewriting

\begin{equation*}
\begin{split}
\mathcal{F}_\mathrm{AQS}(k)&=\text{Tr}[\rho_{\text{a}}(k)|\chi(k) \rangle \langle \chi(k) |] \\&=\text{Tr}[\rho_{\text{a}}(k)W^k|\chi(0) \rangle \langle \chi(0)|(W^{\dagger})^k]\\&=\text{Tr}[(W^{\dagger})^k \rho_{\text{a}}(k)W^k|\chi(0) \rangle \langle \chi(0)|],
\end{split}
\end{equation*}
\\
\noindent which is the overlap between the initial state, $|\chi(0) \rangle \langle \chi(0)|$, and the final state evolved backwards in time without errors, $(W^{\dagger})^k \rho_{\text{a}}(k)W^k$.

\begin{figure*}[!t]
\includegraphics[width=170mm]{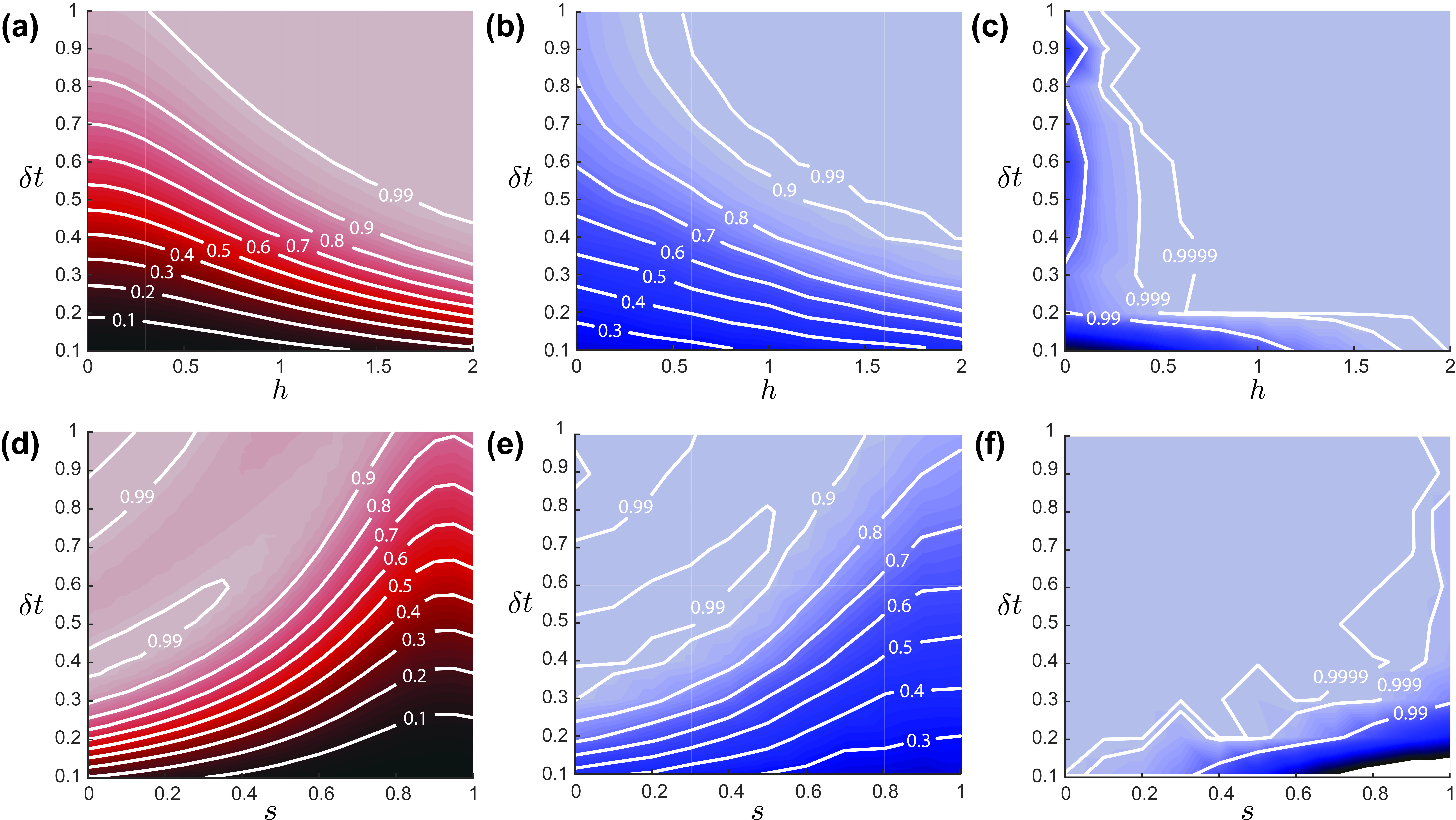}
\centering
\caption{(a) Hilbert-Schmidt distance $||W-\mathcal{I}||$ between the time propagator $W$ and the identity $\mathcal{I}$, for a Transverse Ising model with $s=1$ and time step $\delta t$ (see Eq. 2a of the main text), normalized so the largest distance equals one within the parameter range shown. (b) Mean and (c) maximum fidelity observed for 500 EVO searches with $N_{\phi}=20$. (e-f) Same quantities for the Lipkin-Meshkov-Glick model (Eq. 2b of the main text). As the Hilbert-Schmidt distance increases, the likelihood of finding high fidelity EVO controls grows and then plateaus. Note that even in the near-identity region, we can mostly find useful controls by doing enough searches with different seeds.}
\end{figure*} 

\section{Optimal Control}
\noindent Introductions to Optimal Control can be found in the literature \cite{Brif2010}. In the generic formulation, one starts with a control Hamiltonian $H_{\text{c}} (t) = H_{0} + \sum_{j} b_{j}(t) H_{j}$, chosen so it can generate all possible unitary maps and renders the system controllable. The control waveforms are coarse grained in time, $\{b_{j}(t)\} \rightarrow \{b_{j}(t_{i})\}$, to yield a discrete set of control variables. Given a target unitary $W$ acting in $\mathscr{H}$, one can search for a set $\{b_{j}(t_{i})\}$ that minimizes the Hilbert-Schmidt distance $||W - U(\{b_{j}(t_{i})\})||$, where $ U(\{b_{j}(t_{i})\})$ is the map generated by $H_{\text{c}}(t)$ evolving with a particular set of control variables during the time $T$. If the overall phase of $W$ is unimportant, one can instead maximize the fidelity $\mathcal{F} \{b_{j}(t_{i})\} = | \text{Tr} [ W^{\dagger}  U(\{b_{j}(t_{i})\}) ] |^2 / d^{2}$.        

\subsection{Conventional Control}
\noindent In a variation on the generic formulation, we choose as our control variables a set of $3N_{\phi}$ piecewise constant phases $\{ \phi_{i}^{\text{rf}x}, \phi_{i}^{\text{rf}y} \phi_{i}^{\mu\text{w}} \} = \{ \vec{\phi_{i}} \}$, $1\leq i \leq N_{\phi}$, for the rf and $\mu$w fields. Given a target $W$ and controls $\{ \vec{\phi_{i}} \}$, it is then straightforward to numerically integrate the Schr{\"o}dinger equation to find the resulting transformation $U\{ \vec{\phi_{i}} \}$ and the fidelity $\mathcal{F}(\{\vec{\phi}_{i}\})=|\text{Tr}[W^{\dagger}U(\{\vec{\phi}_{i})\})]|^2/{d^2}$. This allows us to start from a random guess for the $\{ \vec{\phi_{i}} \}$ and use a gradient ascent algorithm \cite{Khaneja2005} to converge on a local maximum of $\mathcal{F}(\{\vec{\phi}_{i}\})$.  The search landscape for this objective function is known to be benign, as long as one allows a sufficient number of phase steps for the system to be controllable.

This protocol is what we refer to as ``Conventional Control" in the main text.  In practical terms, the search complexity is mostly determined by the number of control phases.  When $N_{\phi}$ is sufficiently large, a set of controls that exceed a minimum acceptable fidelity of $\mathcal{F}=0.999$ can usually be found with a single search starting from a single seed.  This typically takes a little over an hour on a desktop computer. Increasing the minimum acceptable fidelity lengthens the search time considerably. We can speed up performance using a dedicated supercomputing core, but a more substantial benefit can be achieved by running multiple searches with different seeds in parallel on multiple cores in a cluster.  If we choose $N_{\phi}=150$, a single search starting from a single random seed will reliably find controls that can implement $W$ to near-arbitrary precision. As $N_{\phi}$ is reduced, sub-optimal traps start to appear in the search landscape, and an increasing number of searches using different random seeds are required to find a high performing set of control phases. Eventually, as $N_{\phi}$ is reduced below some threshold value, Conventional Control will be unable to find a useful set of control phases even after thousands of searches and seeds.

\subsection{EigenValue Only Control}
\noindent For the purpose of quantum simulation one must choose a map $V$ from system to simulator, and because our simulator is fully controllable there are in principle no restrictions on how this is done.  As described in the main text, EigenValue Only (EVO) Control takes advantage of this by optimizing an objective function $\mathcal{F_{\text{EVO}}}(\{\vec{\phi_{i}}\})=|\text{Tr}[W^{\dagger} V U(\{\vec{\phi_{i}}\})V^{\dagger}]|^{2}/d^{2}$.  Our experience shows that the search landscape for EVO Control is less benign than for Conventional Control, and as a result, searches regularly become stuck in traps in the search landscape and fail to converge on usable controls. 
We do not have a good understanding why this should be the case.  However, it is notable that the proliferation of suboptimal traps for the shortest controls is common to both Conventional and EVO Control, and that in both cases the problem gradually disappears as the control time is increased. This echoes findings of Rabitz and co-workers in \cite{Chakrabati2008, Moore2011}, as well as our group in \cite{Anderson2015}.  Regardless, given a sufficient number of initial seeds, EVO optimization will reliably find controls for which $U\{\vec{\phi_{i}}\}$ closely approximates $W$, and will do so with many fewer control phases than required for Conventional Control. Optimizing over fewer phases will sharply reduce the time required to converge on good controls, or to find that the search is stuck in a trap and needs restarting from a different seed. Thus, while EVO Control may need to explore a larger number of initial seeds, we have found that the total computational effort ends up being comparable to Conventional Control. Furthermore, the search for EVO controls is well suited for running seeds in parallel on many cores in a cluster. Out of 2000 initial seeds, we regularly find hundreds of suitable EVO controls that reach $\mathcal{F_{\text{EVO}}} \geq 0.99999$ with only $N_{\phi}=20$ phase steps. In cases where the target $W$ is far from the identity, we have occasionally found waveforms with similar fidelity for as few as  $N_{\phi}=10$ phase steps. 

\subsection{Robust Optimal Control}
\noindent In practice, a control Hamiltonian $H_{c}(t)$ will depend on several parameters $\Theta = \{\theta_{i}\}$ that are imperfectly known. If so, one can search for robust control waveforms by maximizing the average fidelity $\bar{\mathcal{F}} = \int_{\Theta} \mathcal{P}(\Theta)\mathcal{F}(\Theta) d\Theta$, where $\mathcal{P}(\Theta)$ is the probability that the parameters take on the values $\Theta$, and $\mathcal{F}(\Theta)$ is the corresponding fidelity. When searching for robust control waveforms, we have found that our dominant source of uncertainty is variation in the bias Larmor frequency $\Omega_0$ due to spatial inhomogeneity of the bias field strength $B_{0}$, and that maximizing a two-point average $\tilde{\mathcal{F}} = [\mathcal{F}(B_{0} + \delta B) + \mathcal{F}(B_{0} - \delta B)]/2$ is sufficient for good performance \cite{Anderson2015}.  This approach works well for both Conventional and EVO Control, and we have used it consistently for every unitary map used in the experiments described in the main text.

\subsection{Optimal Control Near the Identity}
\noindent We have found empirically that the performance of EVO control depends critically on the Hilbert-Schmidt distance between the target map $W$ and the identity $\mathcal{I}$. To illustrate this point, Fig. 2 shows $||W-\mathcal{I}||$, along with the mean and maximum fidelity resulting from 500 searches for EVO controls with $N_{\phi}=20$.  Figures 2a-c (top row) are for a Transverse Ising model $H_{\text{TI}}$ with varying $h$ and time step $\delta t$. while Figs. 2c-e (bottom row) are for a Lipkin-Meshkov-Glick model $H_\text{LMG}$ with varying $s$ and $\delta_t$. Both examples show strong correlation: when the distance $||W-\mathcal{I}||$ is small, both the mean and maximum fidelities are low; when the distance is large, both are high. When the number of phase steps is increased to $N_{\phi}=150$, high fidelity controls can be found for any $W$.